\documentclass[aps, prd, twocolumn, nofootinbib, floatfix, superscriptaddress]{revtex4-1}

\usepackage{graphicx}
\usepackage{dcolumn}
\usepackage{bm}
\usepackage{subfigure}
\usepackage{epsfig}
\usepackage{amsmath}
\usepackage{amsfonts}
\usepackage{graphicx}
\usepackage{amssymb}
\usepackage{amssymb,amsmath,amsfonts}
\usepackage{xfrac}
\usepackage{color}
\usepackage{tikz}
\usepackage[T1]{fontenc}
\usepackage[utf8]{inputenc}

\usepackage{tcolorbox}
\usepackage{hyperref}
\hypersetup{colorlinks, citecolor=bluscuro, linkcolor=black, urlcolor=bluscuro}
\definecolor{rossos}{cmyk}{0,1,1,0.55}
\definecolor{bluscuro}{rgb}{0.15, 0.2, .85}
\definecolor{bluchiaro}{cmyk}{1,.3,0.,0.1}

\let\oldsqrt\sqrt
\def\sqrt{\mathpalette\DHLhksqrt}
\def\DHLhksqrt#1#2{%
\setbox0=\hbox{$#1\oldsqrt{#2\,}$}\dimen0=\ht0
\advance\dimen0-0.2\ht0
\setbox2=\hbox{\vrule height\ht0 depth -\dimen0}%
{\box0\lower0.4pt\box2}}

\newcommand{\boldmathsymbol}[1]{{\ensuremath{\boldsymbol{#1}}}}

\newcommand{\MeV}{\mathrm{MeV}}

\newcommand{\beq}{\begin{equation}}
\newcommand{\eeq}{\end{equation}}
\newcommand{\bea}{\begin{equation}\begin{aligned}}
\newcommand{\eea}{\end{aligned}\end{equation}}

\newlength{\wsingfig}
\newlength{\wdblefig}
\newlength{\wquadfig}
\newlength{\wtriplefig}
\newcommand{\Eq}[1]{Eq.~(\ref{#1})}

\newcommand{\Fig}[1]{Fig.~{\ref{#1}}}

\newcommand{\be}{\begin{equation}}

\def\apj{ApJ}
\def\mnras{MNRAS}

\def\prd{Phys. Rev. D}
\def\aap{A\&A}                   
\def\apjl{ApJ}                   

\def\prl{PRL}

\begin{document}

\title{Primordial magnetic field generation via primordial black hole disks}

\author{Theodoros Papanikolaou}
\email{papaniko@noa.gr}
 \affiliation{National Observatory of Athens, Lofos Nymfon, 11852 Athens, 
Greece}
\author{Konstantinos N. Gourgouliatos}%
 \email{kngourg@upatras.gr}
\affiliation{Department of Physics, University of Patras, Patras, Rio, 26504, Greece}%


\begin{abstract} 
Large scale primordial magnetic fields (PMFs) threading the intergalactic medium are observed ubiquitously in the Universe playing a key role in the cosmic evolution. Their origin is still debated constituting a very active field of research. In the present article, we propose a novel natural ab initio mechanism for the origin of such PMFs through the portal of supermassive primordial black holes (PBHs) forming between the Big Bang Nucleosynthesis and the recombination era. In particular, by considering PBHs furnished with a locally isothermal disk we study the generation of a Biermann battery induced seed magnetic field (MF) due to the vortexlike motion of the primordial plasma around the black hole. Finally, by considering monochromatic PBH mass distributions and deriving the relevant MF power spectrum we make a conservative estimate for the seed PMF in intergalactic scales and at redshift $z=30$, when typical galaxies are considered to form, which reads as $B\simeq 10^{-30}\mathrm{G}\left(\frac{\ell_\mathrm{R}}{10^6}\right)^2\left(\frac{M_\mathrm{PBH}}{10^{14}M_\odot}\right)^{5/2}$, where $M_\mathrm{PBH}$ is the PBH mass and $\ell_\mathrm{R}\equiv R_\mathrm{d}/R_\mathrm{ISCO}$, is the ratio of the radius of the disk, $R_\mathrm{d}$ over the radius of the innermost stable circular orbit, $R_\mathrm{ISCO}$. Interestingly enough, by requiring to seed a PMF of the order of $10^{-30}\mathrm{G}$ necessary to give rise to a present day $10^{-18}\mathrm{G}$ in intergalactic scales, we find a lower bound on the PBH mass within the range $[10^{10}- 10^{16}]M_\odot$ depending on the radius of the PBH disk.

\end{abstract}

\maketitle


{\bf{Introduction.}} Magnetic fields (MFs) are ubiquitously present in the Universe affecting unavoidably its dynamical evolution at all scales being accompanied with a very rich phenomenology. In particular, magnetic fields can play a key role in the process of particle acceleration through the intergalactic medium~\cite{Bagchi:2002vf} as well as on the propagation of cosmic rays through galaxies and clusters of galaxies~\cite{Strong:2007nh}. They can influence as well the dynamical evolution of the primordial plasma in the early Universe affecting drastically the Universe's thermal state between inflation and recombination. Very interestingly, they can have a relevant impact on CMB anisotropies~\cite{Barrow:1997mj,Subramanian:2006xs} as well as on the physics of Big Bang Nucleosynthesis (BBN)~\cite{Grasso:1996kk}.

However, their dynamical behavior, amplification and above all, their generation mechanism are still not clear, making the area of primordial magnetic field (PMF) astronomy one of the most active and challenging research directions in cosmology. Among their generation mechanisms one should discriminate between cosmological and astrophysical ones. Regarding the cosmological MF generation mechanisms, it has been proposed that seed MFs may arise from  phase transitions in the early Universe~\cite{1988PhRvD..37.2743T,1989ApJ...344L..49Q}, magnetic helicities~\cite{Joyce:1997uy,Giovannini:1997eg}, magnetic monopoles~\cite{Long:2015cza} and primordial scalar~\cite{Ichiki:2006cd,Naoz:2013wla} and vector perturbations~\cite{Banerjee:2003xk,Durrer:2006pc} magnetising the Universe mainly on large-scales. With respect to the astrophysical mechanisms, these can seed PMFs through battery-like mechanisms~\cite{1950ZNatA...5...65B} processing among others in intergalactic shock waves, stellar winds and supernova explosions~\cite{Widrow:2002ud,Hanayama:2005hd} which later can be amplified through turbulence, dynamo processes, differential rotation and instabilities of various types \cite{2006MNRAS.366.1437S,2012PhRvL.108c5002B,2015ApJ...808...65F,2016ApJ...817..127B,2021NatAs...5..268D}.

As regards now the order of magnitude for the amplitude of the MFs in the Universe, we observe MFs with present day amplitudes up to $10^{-7}\mathrm{G}$ on galactic scales~\cite{2004NewAR..48..763V,2011ApJ...728...97V,2019A&A...622A..16O} while on intergalactic scales, there is strong evidence for a pre-galactic seed magnetic field of the order of $10^{-16}\mathrm{G}$ based on observations of high-energy TeV photons emitted by distant blazars~\cite{2010Sci...328...73N}. More recent studies give a reduced MF field strength of the order of $10^{-18}\mathrm{G}$~\cite{2011ApJ...733L..21D}. 
In smaller scales, the intensity of the magnetic field is strongly affected by the presence of interstellar gas and the proximity to stars. For instance in the vicinity of the Earth the interplanetary magnetic field is $10^{-4}$G \cite{2013LRSP...10....5O}.

Accounting therefore for the very rich phenomenology and the numerous generation mechanisms of PMFs we propose in this article a natural generation mechanism of seed PMFs through the portal of primordial black holes (PBHs)~\cite{1967SvA....10..602Z, Carr:1974nx,1975ApJ...201....1C}. These ultra-compact objects can form in the very early Universe well before star formation due to the gravitational collapse of enhanced cosmological perturbations and they can naturally account for a fraction or even the totality of dark matter~\cite{Chapline:1975ojl,Clesse:2017bsw} explaining as well as large-scale structure formation through Poisson fluctuations~\cite{Meszaros:1975ef,Afshordi:2003zb,1984MNRAS.206..315C, Bean:2002kx}. Moreover,they can provide the seeds of the supermassive black holes residing in galactic centres~\cite{1984MNRAS.206..315C, Bean:2002kx,DeLuca:2022bjs} while at the same time they are associated with a plethora of gravitational-wave (GW) signals~\cite{Nakamura:1997sm, Ioka:1998nz, 
Eroshenko:2016hmn,Domenech:2021ztg,Papanikolaou:2020qtd}. 

In this work, we consider the impact of PBHs on the generation of a seed primordial magnetic field. In particular, we consider supermassive PBHs with masses $M_\mathrm{PBH}\in[10^7,10^{18}]M_\odot$ forming during the Hot Big Bang radiation-dominated era after BBN and before recombination era and which are furnished with a battery-induced magnetic field due to the presence of a disk around them whose plasma exhibits a vortexlike motion~\cite{2018MNRAS.479..315S}.


{\bf{The Biermann battery mechanism.}} In general, the creation of an initial seed magnetic field requires the relative motion between negative and positive charges. In astrophysical and cosmological contexts, this role can be played by the Biermann battery mechanism that leads to the spontaneous generation of a magnetic field \cite{1950ZNatA...5...65B,1998ApJ...508..859C,2006ApJ...652.1451C,2018MNRAS.473..721C}. The Biermann battery operates in a system where the charged species have different masses while the energy density and temperature gradients should not be parallel to each other \cite{1983MNRAS.204.1025B}. At the end, one is met with the following term in the magnetic field induction equation 
\begin{eqnarray}
    \partial_t {\bf B}= \nabla \left({\bf u} \times {\bf B}\right) -\frac{c k_B}{e} \frac{\nabla \rho \times \nabla T}{\rho}\,, 
    \label{eq:induction}
\end{eqnarray}

where ${\bf B}$ is the magnetic field, ${\bf u}$ is the plasma velocity, $c$ is the speed of light, $e$ is the electron charge, $k_B$ is the Boltzmann constant, $\rho$ is the plasma energy density and $T$ the temperature.

%
%
The physical interpretation of the Biermann battery is related to the fact that if the energy density gradient is not parallel to the temperature gradient, an electric current is generated. The first term on the right hand side of the induction equation (\ref{eq:induction}), usually called convective term, involving the fluid velocity, may amplify an existing magnetic field through some type of dynamo action, while the second one generates the magnetic field without the need of a seed field. 

{\bf{Locally Isothermal Disks.}} 
The mechanism of Biermann battery operates naturally in a disk environment, provided that the temperature and the density are not functions of each other. This requirement rules out the possibility of the battery mechanism operating in isothermal or barotropic disks~\cite{2018MNRAS.479..315S} since in these regimes the temperature is an explicit function of the energy density. Furthermore, accretion disks require some viscous torques which introduce additional highly uncertain parameters in the system \cite{1973A&A....24..337S,1974MNRAS.168..603L}, namely viscosity at cosmological environments, while at the same time, the knowledge in the literature regarding the accretion disk geometry is quite poor~\cite{Piga:2022ysp}\footnote{Regarding the issue of the geometry of the plasma orbiting around the PBH we cannot assume a Bondi-type spherical accretion disk, since in this case all quantities depend on the radial coordinate and it is impossible to have cross-product terms that will activate the Biermann battery. Thus, we rather consider a disk geometry for the plasma rotating around the black hole and generating the Biermann battery induced MF.}. For this reason, we do not include accretion in our analysis.

A viable possibility, without major ad hoc assumptions, or the introduction of highly uncertain parameters, such as viscosity at cosmological environments, is to consider a locally isothermal disk \cite{2010ApJ...724..730D}, where the density and the pressure are related through the following relation:
\begin{eqnarray}\label{eq:local_isothermal}
    p(R,\phi,z)= \rho(R, \phi, z) c_s^2(R)\,,
\end{eqnarray}
where $(R, \phi, z)$ are the cylindrical coordinates. Thus, while the pressure and density are functions of the spatial coordinates, the speed of sound is a function of the cylindrical radius $R$ only and as a consequence the temperature and energy density profiles are not functions of each other, leading in this way to a Biermann battery induced seed PMF as it can be seen by \Eq{eq:induction}.

Regarding the physical justification of \Eq{eq:local_isothermal}, one should stress that the locally isothermal equation of state, namely \Eq{eq:local_isothermal}, can describe quite well a gas that radiates internal energy gained by shocks~\cite{1999ApJ...526.1001L} (here created by the turbulent motion of the primordial plasma expected after BBN and before recombination era~\cite{1974SvA....18..157K,1974SvA....18..300K,Trivedi:2018ejz,RoperPol:2021gjc} when supermassive PBHs, as the ones we focus on, are expected to form) with high efficiency and can be easily implemented as well within radiation relativistic environments~\cite{2006A&A...445..747K}.

At the end, considering that the unperturbed disk is axially symmetric the locally isothermal disk can be described through an exact solution corresponding to the gravitational potential of a point mass located at the origin:
\begin{eqnarray}
    \Phi = -\frac{GM}{\left(R^2+z^2\right)^{1/2}}\,.
\end{eqnarray}
The pressure and the density are related through the expression:
\begin{eqnarray}
    \frac{p}{\rho}= \epsilon^2 \frac{GM}{R}\,,
    \label{eq:pressure_density}
\end{eqnarray}
where $\epsilon^2$ is a constant and the density reads as
\begin{eqnarray}
    \rho(R,z)=f(R) \exp\left(\frac{R-\sqrt{R^2+z^2}}{\epsilon^2\sqrt{R^2+z^2}}\right)\,,
    \label{eq:density}
\end{eqnarray}
while the angular frequency is given by
\begin{eqnarray}
    \Omega = \frac{GM}{R^3} \left[\frac{R}{\sqrt{R^2+z^2}}-\epsilon^2\left(1-\frac{d \ln f}{d \ln R}\right)\right]\,, 
\end{eqnarray}
where $f(R)$ is a profile function for the density. 

As regards the value of the constant $\epsilon^2$ it can be determined from the boundary condition $\frac{p}{\rho}\vert_{R=R_\mathrm{ISCO}} = c^2/3$ where $R_\mathrm{ISCO}\equiv 6GM/c^2$ stands for the radial distance of the innermost stable circular orbit (ISCO), where the plasma is assumed to be relativistic. At the end, one finds that $\epsilon^2=2$.


{\bf{The growth rate of the seed primordial magnetic field.}} To extract now the seed magnetic field generated through the Biermann battery mechanism we can further relate pressure, density and temperature through the equation of state of an ideal gas:
\begin{eqnarray}
    p = \frac{\rho}{\mu m_\mathrm{e}}k_B T\,,
    \label{eq:state}
\end{eqnarray}
where $m_\mathrm{e}$ is the electron mass, which is actually the mass of the lightest species present in the primordial plasma, and $\mu$ the molecular weight. For our numerical applications, we use $\mu=2$.

At this point, let us stress that the ideal gas equation of state (EoS) used above is strictly valid under the assumption that the interactions between the particles of the gas are negligible independently of the thermal status of the particles in the plasma (relativistic or non-relativistic)~\cite{1999phst.book.....P}. However, in our setup, given the fact that we consider supermassive PBHs forming after BBN happening around $1\mathrm{MeV}$, all the particle species except photons and neutrinos have become non-relativistic. In addition, given the fact that the interactions between the particles within the primordial plasma are quite complicated to be modelled analytically, hence requiring high cost numerical simulations, we consider for simplicity as a first approximation that only the lightest species present, namely the electrons, which have the greatest mobility, will contribute mostly in the pressure of the primordial plasma. This assumption is very well justified for the cosmic epoch between BBN and the recombination era since at that time the species mostly present in the primordial plasma are the electrons and protons before they recombine definitely to form hydrogen atoms. Thus, one can write the EoS for such a plasma in terms of an ideal gas EoS~\cite{Kolb:1990vq}. 

One then can combine the ideal gas EoS (\ref{eq:state}) with (\ref{eq:pressure_density}), in order to get the temperature radial profile which can be recast as follows:
\begin{eqnarray}
    T = \frac{2\mu m_\mathrm{e}}{k_B}\frac{GM}{R}\,,
    \label{eq:Temperature}
\end{eqnarray}
from where we see that the temperature is inversely proportional to the axial distance from the central mass. 
Moreover, substituting the value of $R_\mathrm{ISCO}$ in the above relation we obtain a temperature of the order of $T \approx 4 \times 10^{9}K \simeq 0.35\mathrm{MeV}$, which is comparable to the temperature corresponding to electron decoupling ($T_\mathrm{dec,e}\simeq 6
\times 10^{9} \simeq 0.5\mathrm{MeV}$). Thus, the assumption made above for a relativistic sound speed at $R=R_\mathrm{ISCO}$ is very well justified. 

Finally, through equations (\ref{eq:density}) and (\ref{eq:Temperature}) we can evaluate the Biermann battery term of the induction equation (\ref{eq:induction}) which reads as
\begin{eqnarray}
\begin{split}
   \partial_t {\bf B} & =  \frac{c k_B}{e} \frac{\nabla \rho \times \nabla T}{\rho } 
   =  - \frac{c \mu m_\mathrm{e} GM}{e R^2} \frac{zR}{(R^2+z^2)^{3/2}} \hat{\phi}
    \,.
    \end{split}\label{eq:seed_B_Biermann}
\end{eqnarray}
At the end, one obtains a toroidal seed magnetic field that is antisymmetric with respect to the equatorial plane. To get an order of magnitude estimate for the growth rate of this seed magnetic field we compute it at $R=R_\mathrm{ISCO}$ and at a height $z=R$ since at the equatorial plane,i.e. at $z=0$, $\boldmathsymbol{B}=\boldmathsymbol{0}$. Straightforwardly, one can show by virtue of \Eq{eq:seed_B_Biermann} that 

\begin{eqnarray}
   \frac{\partial{\bf B}}{\partial t} = &  \frac{\sqrt{2} c^7 \mu m_\mathrm{e}}{864 e G^2M^2} \simeq  10^{-5}\left(\frac{10^3M_\odot}{M}\right)^2 \hat{\phi}\;{\mathrm{(G/s)}}
    \label{eq:PMF}
\end{eqnarray}
Following that, the linear phase of growth will lead to a magnetic field whose intensity depends on the mass of PBH in which the Biermann battery mechanism operates. 


{\bf{Relevant timescales.}}
This linear growth is expected to saturate when the temperature and/or energy density gradients are smoothed out as it can be seen by \Eq{eq:seed_B_Biermann}. This saturation time is basically the minimum between the electron dissipation time $t_\mathrm{dis}$ and the sound wave diffusion time $t_\mathrm{sound}$ defined as~\cite{2016PhPl...23e6304S}
\beq
t_\mathrm{dis} \equiv (T/\nabla T)/u_\mathrm{th,e},\quad t_\mathrm{sound}\equiv (\rho/\nabla \rho)/c_\mathrm{s},
\eeq
where $u_\mathrm{th,e}$ is the thermal velocity of the electrons given by $\frac{1}{2}k_\mathrm{B}T = \frac{1}{2}m_\mathrm{e}u^2_\mathrm{th,e}$ and $c_\mathrm{s}$ is the sound speed which by virtue of \Eq{eq:state}, \Eq{eq:pressure_density} and \Eq{eq:Temperature} will read as 
\beq\label{eq:c_s}
c_\mathrm{s} (R) = \frac{c}{\sqrt{3}}\sqrt{\frac{R_\mathrm{ISCO}}{R}}.
\eeq

Note that for the computation of the energy density gradient we consider its variation along the $\boldmathsymbol{z}$-axis since only the vertical gradient of $\rho$ contributes to the Biermann battery seed magnetic field when crossed with the radial temperature profile. Taking at the end the ratio between $t_\mathrm{dis}$ and $t_\mathrm{sound}$  one finds that
\beq
\frac{t_\mathrm{dis}}{t_\mathrm{sound}} = \sqrt{2}\frac{(1+\frac{z^2}{R^2})^{3/2}}{\frac{z}{R}}\gg 1,
\eeq
since the ratio $t_\mathrm{dis}/t_\mathrm{sound}$ is an increasing function of both $R$ and $z$. For $z=R=R_\mathrm{ISCO}$ (thick disk limit), one can get an estimate for the minimum of the ratio $t_\mathrm{dis}/t_\mathrm{sound}$ which can be shown to be $(t_\mathrm{dis}/t_\mathrm{sound})\lvert_{R=R_\mathrm{ISCO}} = 4>1 $ while in the thin disk limit, i.e. $\frac{z}{R}\rightarrow 0$, one obtains that $\lim_{\frac{z}{R}\rightarrow 0}(t_\mathrm{dis}/t_\mathrm{sound})=\infty$. Thus, in both the thin and the thick disk limits $\frac{t_\mathrm{dis}}{t_\mathrm{sound}}>1$ and as a consequence the saturation time $t_\mathrm{s}$ can be recast as
\beq\label{eq:t_saturation}
t_\mathrm{s} \simeq t_\mathrm{sound} = \frac{2}{c_\mathrm{s}(R)}\frac{(R^2+z^2)^{3/2}}{zR}.
\eeq

Concerning now the dynamical time it is defined as the time scale of the orbital motion, namely as $t_\mathrm{dyn}\equiv 1/\Omega$, or equivalently as the time needed to establish the vertical hydrostatic equilibrium,
\beq\label{eq:t_dynamical}
t_\mathrm{dyn}\equiv \frac{H_\mathrm{d}}{c_\mathrm{s}(R)}\,.
\eeq
where $H_\mathrm{d}$ is the thickness of the disk.
Taking the ratio now between the saturation and the dynamical time one gets that
\beq
\frac{t_\mathrm{s}}{t_\mathrm{dyn}} = \frac{2\left(1+\frac{z^2}{R^2}\right)^{3/2}}{\frac{z}{R}\frac{H_\mathrm{d}}{R}}>1,
\eeq
which is in general larger than one within both the very thin disk limit, i.e. $\frac{H_\mathrm{d}}{R}, \frac{z}{R}\rightarrow 0$ and the thick disk regime $z,H_\mathrm{d} \simeq R$ giving $\frac{t_\mathrm{s}}{t_\mathrm{dyn}}\rightarrow 2>1$. Thus, the magnetic field will have time to enter the phase of its linear growth after the formation of the disk and the establishment of hydrostatic equilibrium. 

{\bf{The seed primordial magnetic field.}}
We will now estimate the strength of the seed primordial magnetic field by accounting for the random distribution of PBHs which is encapsulated in the comoving number density of PBHs per unit mass, $\mathrm{d}n/\mathrm{d}M$. For the cases of monochromatic and extended PBH mass distributions (MDs), $\mathrm{d}n/\mathrm{d}M$ can be recast as follows:
\beq
\frac{\mathrm{d}n}{\mathrm{d}M}=
\begin{cases}
\frac{\delta(M-M_\mathrm{PBH})}{\bar{r}^3_\mathrm{PBH}}\rm{\;for\;monorchromatic\;PBH\;MDs}\\
\frac{\rho_\mathrm{c}}{M^2}\beta(M)\rm{\;for\;extended\;PBH\;MDs}
\end{cases}
\eeq
where $\bar{r}_\mathrm{PBH}$ is the mean PBH separation scale defined as $\bar{r}_\mathrm{PBH}\equiv \left(M_\mathrm{PBH}/\rho_\mathrm{PBH}\right)^{1/3}$ and $\beta(M)$ is the PBH mass function defined as $\beta(M)\equiv\frac{1}{\rho_\mathrm{c}}\frac{\mathrm{d}\rho_\mathrm{PBH}}{\mathrm{d}\ln M_\mathrm{PBH}}$.

Continuing now to the derivation of the mean seed MF one should derive the respective magnetic field power spectrum $P_B$. Doing so, we firstly account for the effect of cosmic expansion and we write \Eq{eq:PMF} as~\cite{Naoz:2013wla}
\beq\label{eq:dynamical_B_k}
\begin{split}
\frac{\partial\left(a^2 \boldmathsymbol{B}
\right)}{\partial t} &  =  10\left(\frac{M_\odot}{M}\right)^2 \left(\frac{R_\mathrm{ISCO}}{R}\right)^3\frac{\frac{z}{R}}{\left(1+\frac{z^2}{R^2}\right)^{3/2}}\hat{\phi} \\ & 
\Rightarrow \dot{\boldmathsymbol{B}} + 2H\boldmathsymbol{B} = \boldmathsymbol{S},
\end{split}
\eeq
where $\boldmathsymbol{S}= 10\left(\frac{M_\odot}{M}\right)^2 \left(\frac{R_\mathrm{ISCO}}{R}\right)^3\frac{\frac{z}{R}}{\left(1+\frac{z^2}{R^2}\right)^{3/2}}\hat{\phi}$.

Consequently, the dynamical evolution of the Battery induced MF will be governed by \Eq{eq:dynamical_B_k} where one sees the presence of a source term $S$ and a friction term $2HB$ due to the effect of cosmic expansion. Interestingly, \Eq{eq:dynamical_B_k} can be solved analytically and its solution can be recast as 
\beq
\boldmathsymbol{B}(R,z,t) = 10\left(\frac{M_\odot}{M}\right)^2 \left(\frac{R_\mathrm{ISCO}}{R}\right)^3\frac{\frac{z}{R}\left(1-\frac{t_\mathrm{ini}}{t}\right)\hat{\phi}}{\left(1+\frac{z^2}{R^2}\right)^{3/2}}\; (\mathrm{G}),
\eeq
where $t_\mathrm{ini}$ is an initial time. Here $t_\mathrm{ini}$ is chosen to be equal to the dynamical time $t_\mathrm{dyn} = 1/\Omega$ when the disk establishes hydrostatic equilibrium in the vertical direction. Given now that $t_\mathrm{s}\simeq 2 t_\mathrm{dyn}$ in both the thin and thick disk approximations as we showed above, one can find that at the saturation time, $t=t_\mathrm{s}$, the MF will read as:
\beq\label{eq:B_k_t_s}
\boldmathsymbol{B}(R,z,t_\mathrm{s}) = 5\left(\frac{M_\odot}{M}\right)^2 \left(\frac{R_\mathrm{ISCO}}{R}\right)^3\frac{\frac{z}{R}}{\left(1+\frac{z^2}{R^2}\right)^{3/2}}\hat{\phi}\; (\mathrm{G}),
\eeq

One then can derive the Fourier transform of the magnetic field, $\boldmathsymbol{B}_k$ by accounting for the random distributions of PBHs. In particular, if $\boldmathsymbol{x}^\prime$ stands for the position of a PBH and $\boldmathsymbol{x}$ for the position of a mass element of the disk around this PBH, the Fourier transform of the magnetic field \eqref{eq:B_k_t_s} will be given by
\beq\label{eq:B_k}
\boldmathsymbol{B}_k=\int_{M_\mathrm{min}}^{M_\mathrm{max}}\mathrm{d}M \frac{\mathrm{d}n}{\mathrm{d}M}\int \Biggl(\int \boldmathsymbol{B}(\boldmathsymbol{x}-\boldmathsymbol{x}^\prime)\mathrm{d}^3\boldmathsymbol{x}^\prime\Biggr)e^{i\boldmathsymbol{k}\cdot\boldmathsymbol{x}}\mathrm{d}^3\boldmathsymbol{x}.
\eeq

Then, the magnetic power spectrum will be computed as follows:
\beq\label{eq:P_B}
P_B(k) \equiv \frac{\langle \boldmathsymbol{B}_k\boldmathsymbol{B}^{*}_k\rangle}{V_k}, 
\eeq
where $V_k \equiv \frac{4\pi}{3}\left(\frac{2\pi}{k}\right)^3$, while the mean magnetic field strength as a function of the comoving scale will read as
\beq\label{eq:mean_B}
\langle |\boldmathsymbol{B}_\boldmathsymbol{k}|\rangle  \equiv \sqrt{\frac{k^3 P_B(k)}{2\pi^2}}.
\eeq

At the end, applying the conservation of magnetic flux the magnetic field will scale with cosmic expansion as $B\propto a^{-2}$, thus at a redshift $z$ the mean magnetic field strength will be diluted by a factor of $(a_\mathrm{s}/a_z)^2$ with $a_\mathrm{s}$ being the scale factor at the saturation time. Considering also the fact that the temperature of the Universe scales as $T \propto a^{-1}$, the fact that the mass within the cosmological horizon $M_\mathrm{H} \simeq 17M_\odot \left(200\mathrm{MeV}/T\right)^2$ and that $a_0/a = 1+z$ one can recast the mean magnetic field strength at redshift $z$ as
\beq\label{eq:mean_B_z}
\langle |\boldmathsymbol{B}_\boldmathsymbol{k}|\rangle(z) \simeq 6\times 10^{-24}\frac{M_H}{M_\odot}(1+z)^2\langle |\boldmathsymbol{B}_\boldmathsymbol{k}|\rangle_{z_\mathrm{s}},
\eeq
where $z_\mathrm{s}$ stands for the redshift at the saturation time\footnote{Note that in \Eq{eq:mean_B_z} we accounted for the fact that $t_\mathrm{dyn}=t_\mathrm{s}/2$ and that $ t_\mathrm{s} \sim 10 t_\mathrm{H}$ at $z\simeq R_\mathrm{ISCO}$ where $t_\mathrm{s}$ minimizes. $t_\mathrm{H}$ is the Hubble time which for a radiation dominated era reads as $t_\mathrm{H} = 1/(2H)$.}. At this point, it is important to stress that the PBH mass is a fraction of the mass within the cosmological horizon, namely that $M = \gamma M_\mathrm{H}$, with the parameter $\gamma\sim O(0.001-1)$ depending on the amplitude of the collapsing energy density perturbation~\cite{Niemeyer:1997mt,Musco:2008hv,Musco:2012au}.

{\bf{Monochromatic PBH mass distributions.}} 
Let us now compute the  mean magnetic field strength at redshift $z$ for monochromatic PBH mass distributions. After a long but straightforward calculation, one can find that in the thin-disk limit where $H_\mathrm{d}\leq R_\mathrm{ISCO}$ [See Appendix A for the technical details] mean MF strength reads as

\beq\label{eq:B_k_z_monochromatic_main}
\langle |\boldmathsymbol{B}_\boldmathsymbol{k}|\rangle(z) \simeq 10^{-67}\gamma\ell^2_\mathrm{R}\Omega_\mathrm{PBH,f}\frac{H_\mathrm{d}}{1\mathrm{mpc}}\left(\frac{M}{M_\odot}\right)\left(\frac{k}{1\mathrm{Mpc^{-1}}}\right)^3(1+z)^2,
\eeq
with $\Omega_\mathrm{PBH,f}$ being the initial PBH abundance at PBH formation time and $\ell_\mathrm{R}$ being defined as the ratio between the radius of the disk over $R_\mathrm{ISCO}$, i.e. $\ell_\mathrm{R}\equiv R_\mathrm{d}/R_\mathrm{ISCO}$. Regarding the possible values of $\Omega_\mathrm{PBH,f}$ one can show that by avoiding PBH overproduction at matter-radiation equality one gets that $\Omega_\mathrm{PBH,eq}<1 \Rightarrow \Omega_\mathrm{PBH,f} < 10^{-9}\left(\frac{M}{\gamma M_\odot}\right)^{1/2}$. 
Given also, the fact that \Eq{eq:B_k_z_monochromatic_main} is valid in the thin-disk limit, where $H_\mathrm{d}\leq R_\mathrm{ISCO}= 3\times 10^{-10}M/M_\odot (\mathrm{mpc})$ one can impose an upper limit on the mean MF strength reading as follows:
\beq
\langle |\boldmathsymbol{B}_\boldmathsymbol{k}|\rangle(z) \lesssim 10^{-86}\gamma^{1/2}\ell^2_\mathrm{R}\left(\frac{M}{M_\odot}\right)^{5/2}\left(\frac{k}{1\mathrm{Mpc^{-1}}}\right)^3(1+z)^2.
\eeq
Thus, for $z=30$, which is the typical redshift around the reionisation epoch, when the first galaxies are considered to form and for $k=100\mathrm{Mpc^{-1}}$ corresponding to $r=10\mathrm{kpc}$, being the order of magnitude of the intergalactic scales, 
one obtains that
\beq\label{B_k_z_30_k_100_Mpc_minus_1}
B(k=100\mathrm{Mpc^{-1}},z=30) \lesssim 10^{-77}\gamma^{1/2}\ell^2_\mathrm{R}\left(\frac{M}{M_\odot}\right)^{5/2}.
\eeq
Interestingly enough, for $\gamma = 0.1$, being the typical value of the cosmological horizon mass collapsing into a PBH and for $\ell_\mathrm{R}\sim 1000$ which is the typical ratio of the radius of an accretion disk around a black hole over $R_\mathrm{ISCO}$~\cite{McKinney:2012vh} and for $M_\mathrm{PBH}\sim 10^{16} - 10^{17}M_\odot$, which are typical PBH masses forming between BBN and the recombination epoch, one gets that $B(k=100\mathrm{Mpc^{-1}},z=30)\simeq 10^{-32} - 10^{-29}\mathrm{G}$, which is actually the minimum seed MF amplitude needed to reach the present-day average magnetic field of order $10^{-18}\mathrm{G}$ on intergalactic scales due to dynamo/turbulence/instability processes~\cite{Vachaspati:2020blt}.

As a consequence, one can safely argue that extremely supermassive PBHs forming between BBN and recombination and furnished with a disk can naturally seed the PMFs in the Universe.
This remarkable finding points towards a new natural ab initio scenario for the generation of cosmic magnetic fields achieved through the portal of PBHs.

{\bf{Discussion.}} 
A key question in cosmic evolution, is the origin of the ubiquitous large scale magnetic fields threading the intergalactic medium. Among many other scenarios for the generation of the primordial magnetic fields the Biermann battery seems to be one of the most natural mechanisms for the generation of such fields. 

In this article, we propose a novel ab initio mechanism for the generation of PMFs considering a Biermann battery induced magnetic field generated due to the presence of a locally isothermal disk around a PBH. PBHs can naturally form in early Universe under many formation channels [See here~\cite{Carr:2020xqk,Escriva:2022yaf} for reviews in the topic] during the radiation-dominated era before recombination when matter is in form of plasma exhibiting strong vortex like motion. Thus, under these circumstances, one expects naturally the necessary conditions for the formation of such disks giving rise to battery induced MFs.

In particular, by deriving the magnetic field power spectrum from the MF induction equation we made a conservative estimate on the seed PMF on intergalactic scales at redshift $z\sim 30$, which is the typical redshift at the epoch of reionisation when the first galaxies are considered to form. At the end, by setting $\gamma = 0.1$, being the typical fraction of $M_\mathrm{H}$ collapsing to a PBH, the mean MF strength \eqref{B_k_z_30_k_100_Mpc_minus_1} can be recast as
\beq\label{eq:B_k_z=30_intergalactic}
B \simeq 10^{-30}\mathrm{G}\left(\frac{\ell_\mathrm{R}}{10^6}\right)^2\left(\frac{M_\mathrm{PBH}}{10^{14}M_\odot}\right)^{5/2}.
\eeq
Interestingly enough, for typical values of  $\ell_\mathrm{R}\sim 1000$ we found a lower PBH mass bound of the order of  $10^{16} - 10^{17}M_\odot$, which are typical PBH masses forming between BBN and the recombination epoch, so as to produce a seed PMF $B(k=100\mathrm{Mpc^{-1}},z=30)\simeq 10^{-32} - 10^{-29}\mathrm{G}$, which is actually the minimum seed MF amplitude needed to give rise to the present-day average magnetic field of order $10^{-18}\mathrm{G}$~\cite{Vachaspati:2020blt}.  

At this point, we should point out that we are quite conservative in our estimation for the lower bound on the PBH mass. In particular, we took $\ell_\mathrm{R}\sim 1000$ which is actually the lower bound on  $\ell_\mathrm{R}$ derived from numerical studies for accretion disks of supermassive black holes of masses of the order of $10^6-10^9M_\odot$~\cite{McKinney:2012vh}. This number can be orders of magnitude larger reaching values up to $10^{11}$ depending on the accretion rate~\cite{McKinney:2012vh} thus bringing the PBH mass lower bound down to $10^{10}M_\odot$ as it can be checked by \Eq{eq:B_k_z=30_intergalactic}. In addition, we worked within the thin-disk limit where $H_\mathrm{d}/R_\mathrm{ISCO}\leq 1$ in order to get an analytic expression for the magnetic field power spectrum. However, $H_\mathrm{d}/R_\mathrm{ISCO}$ can in general be larger than one, thus increasing the seed magnetic field. Furthermore, we accounted only for monochromatic PBH mass functions. If one accounts as well for extended PBH mass distributions, they expect a higher mean MF strength as pointed out by~\cite{Araya:2020tds}. Relaxing therefore these assumptions one expects to lower many orders of magnitude the PBH mass lower bound; hence to that end our estimates are rather conservative.

We should point out as well here that the above mentioned scenario for the generation of PMFs can be further constrained through the numerous observational signatures of PBHs. Indicatively, we mention here the dynamical effect of a PBH onto an astrophysical system, the role of PBHs on large scale structure formation as well as the impact on the products of PBH evaporation on the spectral shape of the CMB radiation. Least but not least, one should highlight the numerous GW signals associated to PBHs, from PBH merging events up to primordial scalar induced GWs. All these observational footprints of PBHs can potentially shed light on the conditions prevailed in the early Universe and further constrain the above mentioned battery induced MF generation mechanism.

Finally, we need to mention that our work can be further extended by studying the late-time evolution of the seed PMFs taking into account the effect of accretion and the possible dynamo/turbulence/instability amplification through the activation of the convective term in the MF induction equation. Furthermore, it will be interesting to explore the statistical battery-induced MF generated from a extended distribution of PBHs with different masses~\cite{Araya:2020tds} as well as the effect of the backreaction of PMFs to the curvature of the spacetime investigating in this way possible effects of such PMFs on the statistical anisotropies of the matter spectrum in a wide variety of scales~\cite{Saga:2020ics} as well as distinctive GW signatures of the above mentioned PMF generation mechanism.

\begin{acknowledgments}
T.P. acknowledges financial support from the Foundation for Education and European Culture in Greece as well as the 
contribution of the COST Actions  
CA18108 ``Quantum Gravity Phenomenology in the multi-messenger approach''  and 
CA21136 ``Addressing observational tensions in cosmology with systematics and 
fundamental physics (CosmoVerse)''. 

This work is part of the activities of the University of Patras GW group of the LISA consortium.
\end{acknowledgments}

\begin{appendix}
\section{Appendix A: The seed magnetic field for a monochromatic PBH mass distribution}
In this appendix, we extract the magnetic field power spectrum and the mean magnetic field strength in the thin-disk limit where $H_\mathrm{d}\leq R_\mathrm{ISCO}$ accounting for a monochromatic PBH mass function. To begin with, we compute firstly the Fourier component of the magnetic field from \Eq{eq:B_k}. Doing so, we compute firstly the integral $I_1(\boldmathsymbol{x}) = \int \boldmathsymbol{B}(\boldmathsymbol{x}-\boldmathsymbol{x}^\prime)\mathrm{d}^3\boldmathsymbol{x}^\prime$ where $\boldmathsymbol{x}^\prime$ stands for the position of the PBH while $\boldmathsymbol{x}$ denotes the position of the mass element of the disk. To simplify the calculation, we consider that both $\boldmathsymbol{x}^\prime$ and $\boldmathsymbol{x}$ are found in the $yz$ plane, accounting in this way for the $\phi$ symmetry of $\boldmathsymbol{B}$ as imposed by \Eq{eq:B_k_t_s}. Thus, one can recast $I_1(\boldmathsymbol{x})$ as follows
\begin{widetext}
\beq
\begin{split}
I_1(\boldmathsymbol{x}) & = \int \boldmathsymbol{B}(\boldmathsymbol{x}-\boldmathsymbol{x}^\prime)\mathrm{d}^3\boldmathsymbol{x}^\prime \\ & = 20\pi R^3_\mathrm{ISCO}\left(\frac{M_\odot}{M}\right)^2\int_0^{R_\mathrm{H}}\mathrm{d}R^\prime\int_{-R_\mathrm{H}}^{R_\mathrm{H}}\mathrm{d}z^\prime \frac{\frac{|z-z^\prime|}{(R-R^\prime)^4}}{\left[1+\left(\frac{z-z^\prime}{R-R^\prime}\right)^2\right]^{3/2}} \\ & 
= 20\pi R^3_\mathrm{ISCO}\left(\frac{M_\odot}{M}\right)^2\Biggl\{ \frac{2R}{z-R_\mathrm{H}} \Biggl[ \mathrm{Arctanh}\left(\frac{R+\sqrt{R^2+(z-R_\mathrm{H})^2}}{z-R_\mathrm{H}}\right) - \mathrm{Arctanh}\left(\frac{R-R_\mathrm{H}+\sqrt{(R-R_\mathrm{H})^2+(z-R_\mathrm{H})^2}}{z-R_\mathrm{H}}\right) \\ & + \mathrm{Arctanh}\left(\frac{R+\sqrt{R^2+(z+R_\mathrm{H})^2}}{z+R_\mathrm{H}}\right) - \mathrm{Arctanh}\left(\frac{R-R_\mathrm{H}+\sqrt{(R-R_\mathrm{H})^2+(z+R_\mathrm{H})^2}}{z+R_\mathrm{H}}\right) \Biggr]  + 2\ln\left(-\frac{R-R_\mathrm{H}}{2R}\right) - \\ & \Biggl[ \ln\left(\frac{z-R_\mathrm{H}}{R+\sqrt{R^2+(z-R_\mathrm{H})^2}}\right)  - \ln\left(\frac{z-R_\mathrm{H}}{R-R_\mathrm{H}+\sqrt{(R-R_\mathrm{H})^2+(z-R_\mathrm{H})^2}}\right)  + \ln\left(\frac{z+R_\mathrm{H}}{R+\sqrt{R^2+(z+R_\mathrm{H})^2}}\right) \\ &  - \ln\left(\frac{z+R_\mathrm{H}}{R-R_\mathrm{H}+\sqrt{(R-R_\mathrm{H})^2+(z+R_\mathrm{H})^2}}\right) \Biggr] \Biggr\},
\end{split}
\eeq
\end{widetext}
where $R_\mathrm{H}\equiv c H^{-1}$ stands for the cosmological horizon scale. In principle, one should take the integrals over $R^\prime$ and $z^\prime$ from $0$ to $\infty$ and from $-\infty$ to $\infty$. However, this leads to divergences, hence we take the integral over $R^\prime$ from $0$ to $R_\mathrm{H}$ and that over $z^\prime$ from $-R_\mathrm{H}$ up to $R_\mathrm{H}$ since in principle there is one PBH within the cosmological horizon at the time of PBH formation. Then, working in the thin-disk limit where $z/R_\mathrm{ISCO} = z/(3\gamma R_\mathrm{H})\leq 1$, one can show by defining the auxiliary variables $y\equiv R/R_\mathrm{H}$ and $p=z/R_\mathrm{H}$ and keeping terms up to $O(p^2)$ that 
\begin{widetext}
\beq
\begin{split}
I_1(\boldmathsymbol{x}) & \simeq 20\pi R^3_\mathrm{ISCO}\left(\frac{M_\odot}{M}\right)^2 \Biggl(4y \mathrm{Arctanh}\left(y+\sqrt{1+y^2}\right) + 4y\mathrm{Arctanh}\left(1-y-\sqrt{2-2y+y^2}\right) + 2\ln\left(\frac{1-y}{2y}\right)
\\ & + \Biggl\{ \frac{2y+5y^3+2y^5+ 4y^2\sqrt{1+y^2} + 2y^4\sqrt{1+y^2}}{2\left(1+y^2\right)^{3/2}\left(y+\sqrt{1+y^2}\right)^2} 
\\ & + \Bigl[9-27y+35y^2-25y^3+10y^4-2y^5-6\sqrt{2-2y+y^2}+16y\sqrt{2-2y+y^2} -16y^2\sqrt{2-2y+y^2} \\ & + 8y^3\sqrt{2-2y+y^2} - 2y^4\sqrt{2-2y+y^2}\Bigr]/\Bigl[\left(2-2y+y^2\right)^{3/2}\left(-1+y+\sqrt{2-2y+y^2}\right)^2\Bigr] 
\\ & + \Biggl[\frac{y^2}{1+y^2} +\frac{2y+3y^3+3y^2\sqrt{1+y^2}}{\left(1+y^2\right)^{3/2}\left(y+\sqrt{1+y^2}\right)^2}\Biggr]+ 2y\Biggl[ -\frac{1}{\sqrt{1+y^2}} + \frac{1}{2\sqrt{2-2y+y^2}} - \frac{-1-2y^2-2y\sqrt{1+y^2}}{2\left(1+y^2\right)^{3/2}\left(y+\sqrt{1+y^2}\right)^2}
\\ & + \frac{3-4y+2y^2-2\sqrt{2-2y+y^2} + 2y\sqrt{2-2y+y^2}}{4\left(2-2y+y^2\right)^{3/2}\left(-1+y+\sqrt{2-2y+y^2}\right)^2} + 2\mathrm{Arctanh}\left(y+\sqrt{1+y^2}\right)-\mathrm{Arctanh}\left(1-y-\sqrt{2-2y+y^2}\right)
\Biggr]
\\ & -2y\Biggl[-3\frac{3-4y+2y^2-2\sqrt{2-2y+y^2} + 2y\sqrt{2-2y+y^2}}{4\left(2-2y+y^2\right)^{3/2}\left(-1+y+\sqrt{2-2y+y^2}\right)^2} - \mathrm{Arcthan}\left(1-y-\sqrt{2-2y+y^2}\right)\Biggr]
\Biggr\}p^2
\Biggr).
\end{split}
\eeq
\end{widetext}

At the end, by Fourier transforming $\boldmathsymbol{B}$ according to \Eq{eq:B_k} we integrate $I_1$ over $y$ from $3\gamma$ up to $3\gamma\ell_\mathrm{R}$ since $R_\mathrm{ISCO}=3\gamma R_\mathrm{H}\leq R\leq \ell_\mathrm{R}R_\mathrm{ISCO}= 3\gamma \ell_\mathrm{R} R_\mathrm{H} $ and over $p$ from $-H_\mathrm{d}/R_\mathrm{H}$ up to $H_\mathrm{d}/R_\mathrm{H}$. Finally, accounting for the fact that $\ell_\mathrm{R}\gg 1$ one gets after the integrations over $y$ and $p$ that $\boldmathsymbol{B}_k$ reads as
\begin{widetext}
\beq\label{eq:B_k_expanded}
\boldmathsymbol{B}_k \simeq 40\pi^3 \gamma^2\ell^2_\mathrm{R} \left(\frac{M_\odot}{M_\mathrm{PBH}}\right)^2\frac{9e^{-ikH_\mathrm{d}}\left(-1+e^{2ikH_\mathrm{d}}\right)}{k}\int_{M_\mathrm{min}}^{M_\mathrm{max}} R^3_\mathrm{ISCO}R^2_\mathrm{H}\frac{\mathrm{d}n}{\mathrm{d}M}\mathrm{d}M.
\eeq
\end{widetext}
In the case of a monocrhomatic PBH mass function one has that $\frac{\mathrm{d}n}{\mathrm{d}M} = \delta(M-M_\mathrm{PBH})/\bar{r}^3_\mathrm{PBH}$ with $\bar{r}_\mathrm{PBH}$ being recast as
\beq\label{PBH mean separation}
\begin{split}
\bar{r}_\mathrm{PBH}(t) & = \left(\frac{M_\mathrm{PBH}}{\rho_\mathrm{PBH}}\right)^{1/3} = \left(\frac{4\gamma\pi \rho_\mathrm{tot,f}H_\mathrm{f}^{-3}/3}{\Omega_\mathrm{PBH}(t)\rho_\mathrm{tot}(t)}\right)^{1/3} \\ &  =\left(\frac{4\pi\gamma}{3\Omega_\mathrm{PBH,f}}\right)^{1/3}c H^{-1}_\mathrm{f}\left(\frac{a}{a_\mathrm{f}}\right).
\end{split}
\eeq
At $t=t_\mathrm{s}$ one gets that $\bar{r}_\mathrm{PBH} = \left(\frac{4\pi\gamma}{3\Omega_\mathrm{PBH,f}}\right)^{1/3}\sqrt{10}R_\mathrm{H}$, where we accounted for the fact that $a\propto t^{1/2}$ in the RD era and that $t_\mathrm{s}\simeq 10 t_\mathrm{H}$ as stated above in the main text. Given now that $R_\mathrm{ISCO}=3\gamma R_\mathrm{H} = 3\gamma cH^{-1} = 3\gamma \times 10^{-19}\mathrm{Mpc}M_\mathrm{H}/ M_\odot$ one obtains from \Eq{eq:B_k_expanded} that 
\beq
\boldmathsymbol{B}_k \simeq \Omega_\mathrm{PBH,f}\gamma^2\ell^2_\mathrm{R}\frac{e^{-ikH_\mathrm{d}}\left(-1+e^{2ikH_\mathrm{d}}\right)}{k}10^{-33}\mathrm{Mpc}^2.
\eeq

Then, from \Eq{eq:P_B}, \Eq{eq:mean_B} and \Eq{eq:mean_B_z} it straightforward to show that

\vspace{2cm}

\begin{widetext}
\beq\label{eq:B_k_z_monochromatic_full}
\langle |\boldmathsymbol{B}_\boldmathsymbol{k}|\rangle(z) \simeq 10^{-58}\gamma \ell^2_\mathrm{R}\Omega_\mathrm{PBH,f}\left(\frac{M_\mathrm{PBH}}{M_\odot}\right)(1+z)^2\left(\frac{k}{1\mathrm{Mpc^{-1}}}\right)^2\sin\left(kH_\mathrm{d}\right).
\eeq
\end{widetext}

At this point, we should point out that the scales one can probe are those larger than the PBH mean separation in order not to probe the granularity of the PBH energy density fluctuations entering potential the non-linear regime. Thus, one can introduce a UV cut-off $k_\mathrm{UV}$ defined as

\beq
k_\mathrm{UV}\equiv 2\pi/\bar{r}_\mathrm{PBH} = \frac{2\pi}{R_\mathrm{H,f}} \left(\frac{3\Omega_\mathrm{PBH,f}}{4\pi\gamma}\right)^{1/3}\frac{a_\mathrm{f}}{a},
\eeq
where in the last equality we used \Eq{PBH mean separation}. At the end, since $k\leq k_\mathrm{UV}$ and $H_\mathrm{d}/R_\mathrm{ISCO}\leq 1$, one has that 

\beq
kH_\mathrm{d}<2\pi/\bar{r}_\mathrm{PBH} = 6\pi \gamma \frac{H_\mathrm{d}}{R_\mathrm{ISCO}} \left(\frac{3\Omega_\mathrm{PBH,f}}{4\pi\gamma}\right)^{1/3}\ll 1,
\eeq
where we accounted for the fact that $R\propto a$ due to cosmic expansion and that $R_\mathrm{ISCO}=3\gamma R_\mathrm{H}$. Thus, $\sin\left(kH_\mathrm{d}\right)\simeq k H_\mathrm{d}$ and one obtains from \Eq{eq:B_k_z_monochromatic_full} that 

\beq\label{eq:B_k_z_monochromatic}
\langle |\boldmathsymbol{B}_\boldmathsymbol{k}|\rangle(z) \simeq 10^{-67}\gamma \ell^2_\mathrm{R}\Omega_\mathrm{PBH,f}\frac{H_\mathrm{d}}{1\mathrm{mpc}}\left(\frac{M_\mathrm{PBH}}{M_\odot}\right)\left(\frac{k}{1\mathrm{Mpc^{-1}}}\right)^3(1+z)^2.
\eeq

\end{appendix}

\bibliography{PBH}

\end{document}